\documentclass[letter]{aa}
\usepackage{graphicx}
\usepackage[varg]{txfonts}
\usepackage{natbib}
\bibpunct{(}{)}{;}{a}{}{,} 
\begin{document}

\title{
Multi-epoch VLTI-PIONIER imaging of the supergiant V766~Cen
\thanks{Based on observations made with the VLT Interferometer
at Paranal Observatory under programme IDs 092.D-0096, 092.C-0312, 
and 097.D-0286.}
\thanks{Olivier Chesneau was PI of the program 092.D-0096. 
He unfortunately passed away before seeing the results coming out of it.
This letter may serve as a posthumous tribute to his 
inspiring work on this source.}
}
\subtitle{Image of the close companion in front of the primary}
\titlerunning{VLTI-PIONIER imaging of V766~Cen and its close companion}
\author{
M.~Wittkowski\inst{1}\and
F.~J.~Abell\'an\inst{2}\and
B.~Arroyo-Torres\inst{3,4}\and
A.~Chiavassa\inst{5}\and
J.~C.~Guirado\inst{2,6}\and
J.~M.~Marcaide\inst{2}\and
A.~Alberdi\inst{3}\and
W.~J.~de~Wit\inst{7}\and
K.-H.~Hofmann\inst{8}\and
A.~Meilland\inst{5}\and
F.~Millour\inst{5}\and
S.~Mohamed\inst{9,10,11}\and
J. Sanchez-Bermudez\inst{12}
}
\institute{
European Southern Observatory, Karl-Schwarzschild-Str. 2,
85748 Garching bei M\"unchen, Germany,
\email{mwittkow@eso.org}
\and
Dpt. Astronomia i Astrof\' isica, Universitat de Val\`encia,
C/Dr. Moliner 50, 46100, Burjassot, Spain
\and
Instituto de Astrof\'isica de Andaluc\'ia (IAA-CSIC),
Glorieta de la Astronom\'ia S/N, 18008, Granada, Spain
\and
Centro Astron\'omico Hispano Alem\'an, Calar Alto, (CSIC-MPG),
Sierra de los Filabres, E-04550 Gergal, Spain
\and
Universit\'e C\^ote d’Azur, Observatoire de la C\^ote d’Azur, CNRS, 
Lagrange, CS 34229, 06304 Nice Cedex 4, France
\and
Observatori Astron\`omic, Universidad de Val\`encia, 46980 Paterna, Spain
\and
European Southern Observatory, Casilla 19001, Santiago 19, Chile
\and
Max-Planck-Institut f\"ur Radioastronomie, Auf dem H\"ugel 69, 53121 Bonn, 
Germany
\and
South African Astronomical Observatory, PO Box 9, Observatory 7935, 
South Africa
\and
Astronomy Department, University of Cape Town, 7701, Rondebosch, South Africa
\and
National Institute for Theoretical Physics, Private Bag X1, Matieland, 7602, 
South Africa
\and
Max-Planck-Institut f\"ur Astronomie, K\"onigstuhl 17, 69117 Heidelberg, 
Germany
}
\date{Received \dots; accepted \dots}
\abstract{The star V766 Cen (=HR 5171A) was originally classified as a 
yellow hypergiant but lately found to more likely be a 27--36\,$M_\odot$ 
red supergiant (RSG).
Recent observations indicated a close eclipsing companion 
in the contact or common-envelope phase.}
{Here, we aim at imaging observations of V766 Cen 
to confirm the presence of the close companion.
}
{We used near-infrared $H$-band aperture synthesis imaging at three 
epochs in 2014, 2016, and 2017, employing the PIONIER instrument at
the Very Large Telescope Interferometer (VLTI).}
{The visibility data indicate a mean Rosseland angular diameter of
4.1$\pm$0.8 mas, corresponding to a radius of 1575$\pm$400\,$R_\odot$. 
The data show an extended shell (MOLsphere) of about 2.5 times 
the Rosseland diameter, which contributes about 30\% of the $H$-band flux.
The reconstructed images at the 2014 epoch show a complex elongated
structure within the photospheric disk with a contrast of about 10\%.
The second and third epochs show qualitatively and 
quantitatively different structures with a single very bright 
and narrow feature and high contrasts of 20--30\%. This
feature is located toward the south-western limb of the photospheric 
stellar disk.
We estimate an angular size of the feature 
of 1.7$\pm$0.3 mas, corresponding to a radius of 
650$\pm$150\,$R_\odot$, and giving a radius ratio of 0.42$^{+0.35}_{-0.10}$ 
compared to the primary stellar disk.}
{We interpret the images at the 2016 and 2017 epochs as showing the 
close companion, or a common envelope toward the companion, 
in front of the primary. At the 2014 epoch, the close companion is behind 
the primary and not visible. Instead, the structure and 
contrast at the 2014 epoch are typical of a single RSG harboring giant 
photospheric convection cells. The companion is most likely a 
cool giant or supergiant star with a mass of 5$^{+15}_{-3}$$M_\odot$.
}
\keywords{
Techniques: interferometric --
stars: massive --
stars: imaging --
supergiants --
binaries: eclipsing --
binaries: close
}
\maketitle
\section{Introduction}
Red supergiants (RSGs) are cool evolved massive stars before their
transition toward core-collapse supernovae (SNe).
Their characterization and location in the Hertzsprung-Russell 
(HR) diagram are important to calibrate stellar evolutionary models of
massive stars and to understand their further evolution toward 
SNe \citep[e.g.,][]{Dessart2013,Groh2013,Groh2014}. 
The majority of massive stars are members of binary systems with a preference
for close pairs
\citep{Podsiadlowski2010,Sana2012}.
Binary interactions have profound implications for 
the late stellar evolution of massive stars toward the different types of SNe
and gamma-ray bursts (GRBs). For example, \citet{Podsiadlowski2017}
and \citet{Menon2017} argued that the progenitor of SN1987A was likely
a blue supergiant that was a member of a close binary system, where the
companion dissolved completely during a common-envelope phase when
the primary was a RSG.

The massive evolved star V766~Cen (=HR~5171~A) was originally classified as a 
yellow hypergiant \citep[YHG,][]{Humphreys1971,vanGenderen1992,deJager1998}. 
It is known to have a wide B0\,Ib companion at a separation 
of 9.7\arcsec. The distance to V766~Cen is well established 
at 3.6$\pm$0.5\,kpc \citep{Chesneau2014}.
\citet[][in the following C14]{Chesneau2014} found evidence
that the primary component itself (HR~5171~A) has an
eclipsing close companion, most likely in a
contact or common-envelope phase.
Based on VLTI-AMBER spectro-interferometry, 
\citet{Wittkowski2017a} reported that V766~Cen is a high-luminosity
($\log L/L_\odot=$\,5.8\,$\pm$\,0.4) source of effective 
temperature 4290\,$\pm$\,760\,K and radius 1490\,$\pm$\,540\,$R_\odot$,
located in the HR diagram close to both the Hayashi and Eddington limits. 
With this location and radius, it is more likely 
a RSG before evolving to a YHG, and consistent with a 40\,$M_\odot$ 
track of current mass 27--36\,$M_\odot$. This mass is consistent
with a system mass of 39$^{+40}_{-22}M_\odot$ and mass ratio 
$q \le 10$ by C14. 

Here, we present near-infrared $H$-band aperture synthesis images of 
V766~Cen with the VLTI-PIONIER instrument at multiple epochs
to detect the close companion by imaging,
and to investigate the surface structure of the primary.
\section{Observations and data reduction}
\label{sec:obs_pionier}
We obtained observations of V766~Cen 
with the PIONIER instrument \citep{LeBouquin2011} of 
the Very Large Telescope Interferometer (VLTI) and its four auxiliary 
telescopes (ATs). 
We took data at three epochs with mean Julian Day 
2456719 (Feb--Mar 2016, duration 11\,d), 2457528 (May--Jul 2016, 
55\,d), and 2457839 (Feb-Apr 2017, 64\,d).
The durations correspond to 0.8\%, 4.2\%, and 4.9\%, respectively, of the 
estimated 1304\,d period of the close companion (C14).
In 2014, the data were dispersed over three spectral channels with
central wavelengths 1.59\,$\mu$m, 1.68\,$\mu$m, 1.77\,$\mu$m
and channel widths of $\sim$\,0.09\,$\mu$m. In 2016 and 2017,
the data were dispersed over six spectral channels with                   
central wavelengths 1.53\,$\mu$m, 1.58\,$\mu$m, 1.63\,$\mu$m, 1.68\,$\mu$m, 
1.72\,$\mu$m, 1.77\,$\mu$m, and widths of $\sim$\,0.05\,$\mu$m.
Observations of V766 Cen were interleaved with observations 
of interferometric calibrators. 
The calibrators were HD~122438 (spectral type K2\,III, angular uniform disk diameter $\Theta_\mathrm{UD}^\mathrm{H}=$1.23\,$\pm$\,0.08\,mas, 
used in 2014),
HD~114837 (F6\,V, 0.78\,$\pm$\,0.06\,mas, 2014), and
HR~5241 (K0\,III, 1.68\,$\pm$\,0.11\,mas, 2016 and 2017).
The angular diameters are from the catalog by \citet{Lafrasse2010}.
Table~\ref{tab:obs_pionier} provides the log of our observations.
Figure~\ref{fig:pionier_uv} shows the $uv$ coverages at each epoch, where
$u$ and $v$ are the spatial coordinates of the baselines projected on sky.
We reduced and calibrated the data with the {\tt pndrs} package
\citep{LeBouquin2011}. Figure~\ref{fig:vis_v766cen} shows all
resulting visibility data of the three epochs together with a model curve
as described in Sect.~\ref{sec:analysis}, and synthetic visibility values 
based on aperture synthesis imaging as described in Sect.~\ref{sec:imaging}.
\section{Data analysis}
\label{sec:analysis}
\begin{table}
\small
\centering
\caption{Fit results including the angular diameters of the photosphere ($\Theta_\mathrm{Ross}$) and the MOLsphere ($\Theta_\mathrm{UD}$), and their flux fractions $f_\mathrm{Ross}$ and $f_\mathrm{UD}$.
\label{tab:fitresults}}
\begin{tabular}{rrrrr}
\hline\hline
Epoch & $\Theta_\mathrm{Ross}$ & $f_\mathrm{Ross}$ & $\Theta_\mathrm{UD}$ & $f_\mathrm{UD}$ \\ 
      & (mas)                  &                   &  (mas)               &                 \\ 
I     & 3.3                    & 0.54              & 5.5                  & 0.44            \\ 
II    & 4.3                    & 0.81              & 12.0                 & 0.17            \\ 
III   & 4.8                    & 0.76              & 10.0                 & 0.23            \\ 
Mean  & 4.1$\pm$0.8            & 0.70$\pm$0.14     & 9.2$\pm$3.3          & 0.28$\pm$0.14   \\ 
\hline
\end{tabular}
\end{table}
The visibility data in Fig.~\ref{fig:vis_v766cen} indicate an overall
spherical stellar disk.
However, deviations from a continuously decreasing visibility in the first lobe
and closure phases different from 0/180\degr
at higher spatial frequencies indicate sub-structure
within the stellar disk.  Changes in the closure phase data 
among the three epochs indicate variability of the structure with time.

Previous observations \citep[C14,][]{Wittkowski2017a}
indicated the presence of an extended molecular layer, also called 
MOLsphere \citep{Tsuji2000}. We used a two-component model for
the overall stellar disk, consisting of a PHOENIX model atmosphere
\citep{Hauschildt1999a} describing the stellar photosphere and a 
uniform disk (UD) describing the MOLsphere. 
We chose a PHOENIX model from the grid of \citet{Arroyo2013}
with parameters close to the values from
\citet{Wittkowski2017a}: mass 20\,$M_\odot$,
effective temperature 3900\,K, surface gravity $\log g$=-0.5,
 and solar metallicity. 
The fit was performed in the same way as in
\citet{Wittkowski2017a}. We treated the flux fractions 
$f_\mathrm{Ross}$ and $f_\mathrm{UD}$ both as free parameters to allow for
an additional over-resolved background component.

Figure~\ref{fig:vis_v766cen} shows our best-fit models
compared to the measured data, showing that the model is successful
in describing the visibility data in the 
first lobe. The contribution of the PHOENIX model alone is plotted to 
illustrate that a single-component model cannot reproduce the 
measured shape of the visibility function.
Table~\ref{tab:fitresults} shows the best-fit
parameters for each epoch. 

Differences among the three epochs may be caused by a 
variability of the overall source structure, or by
systematic effects such as the sparse coverage of visibility points
at low spatial frequencies at epoch I.
We used the averaged values and their standard deviations
as final fit results as listed in Tab.~\ref{tab:fitresults}.
Our value of the Rosseland angular diameter $\Theta_\mathrm{Ross}$ 
of 4.1$\pm$0.8\,mas is consistent with the estimates of 
$\Theta_\mathrm{UD}$=3.4$\pm$0.2\,mas by C14 and
$\Theta_\mathrm{Ross}$=3.9$\pm$1.3\,mas by \citet{Wittkowski2017a}.
\section{Aperture synthesis imaging}
\label{sec:imaging}
\begin{figure*}
\centering
  \includegraphics[width=0.255\hsize]{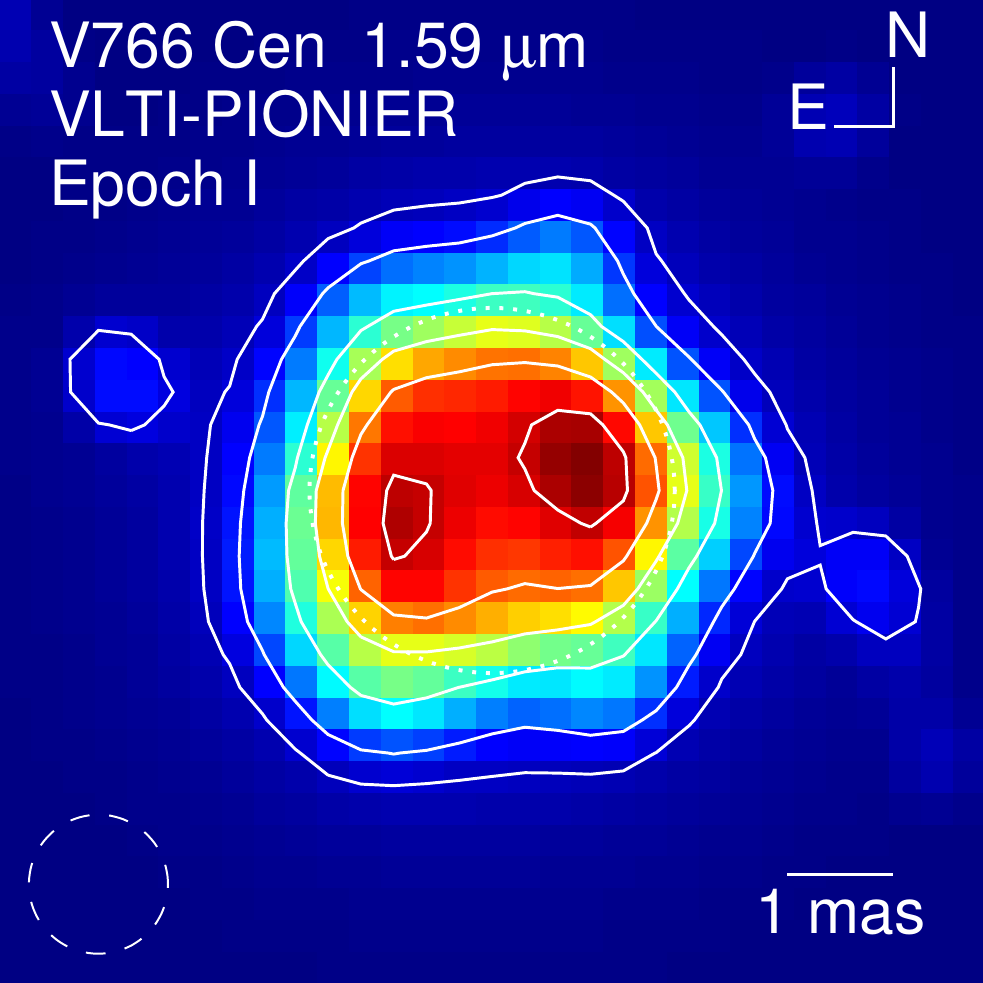}
  \includegraphics[width=0.255\hsize]{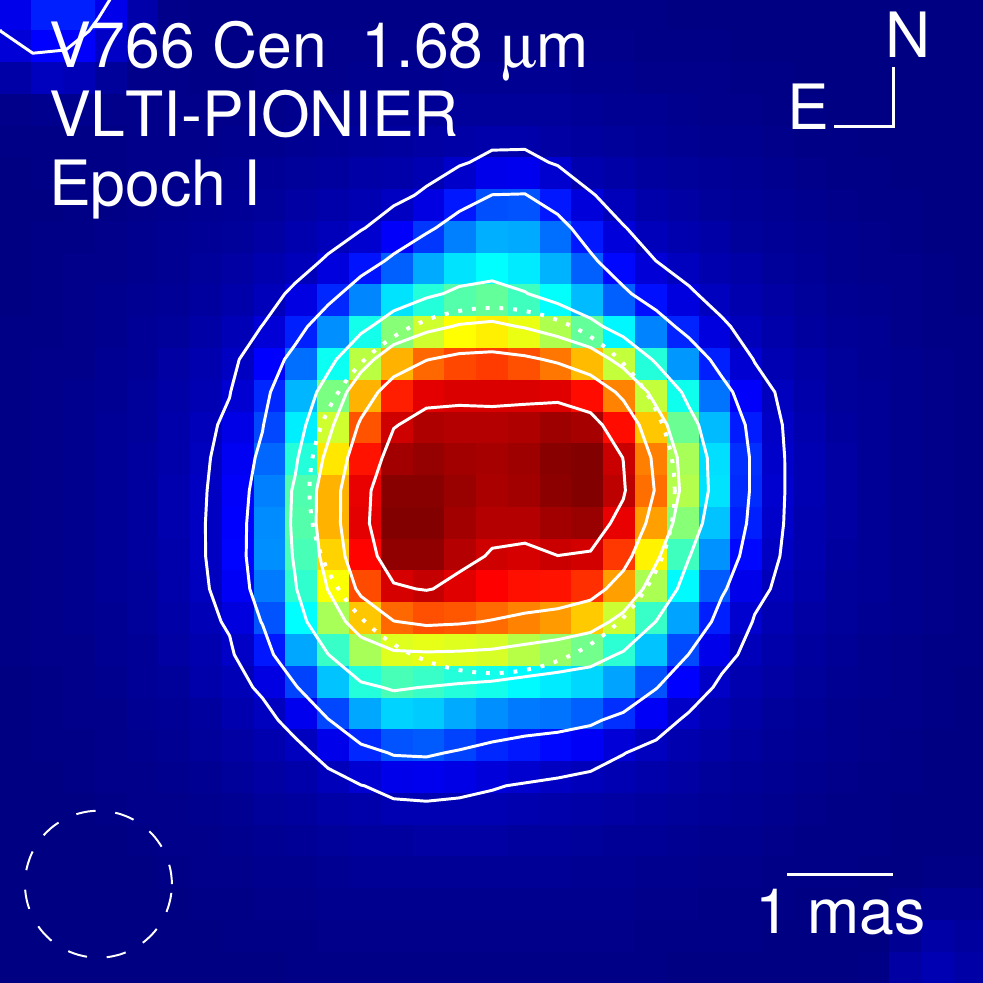}
  \includegraphics[width=0.255\hsize]{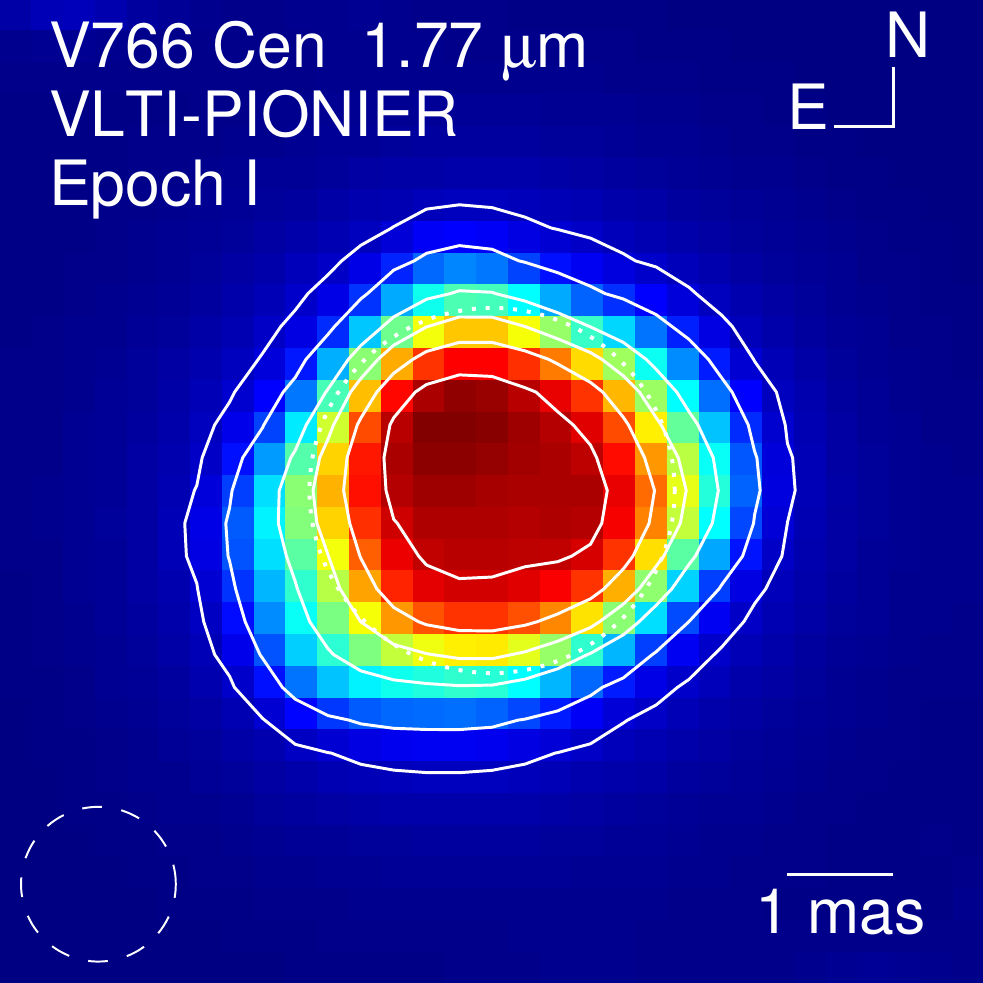}

  \includegraphics[width=0.255\hsize]{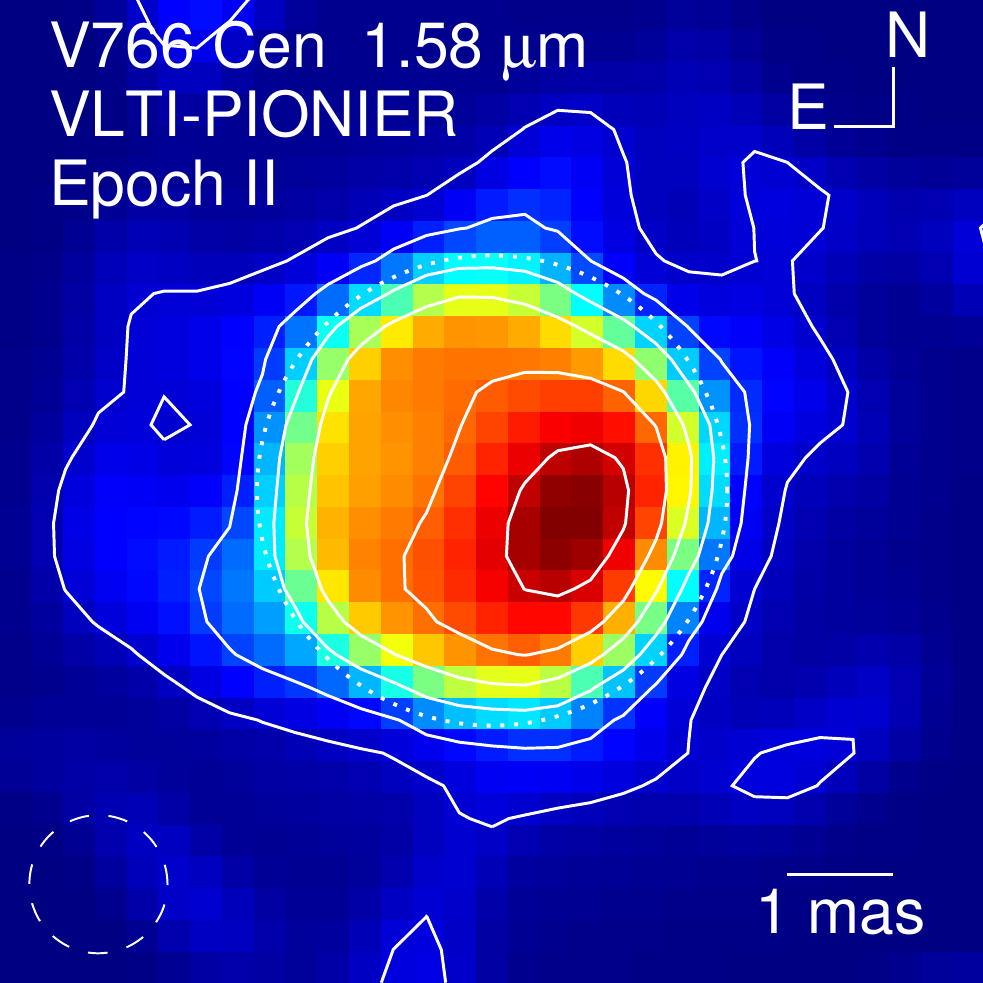}
  \includegraphics[width=0.255\hsize]{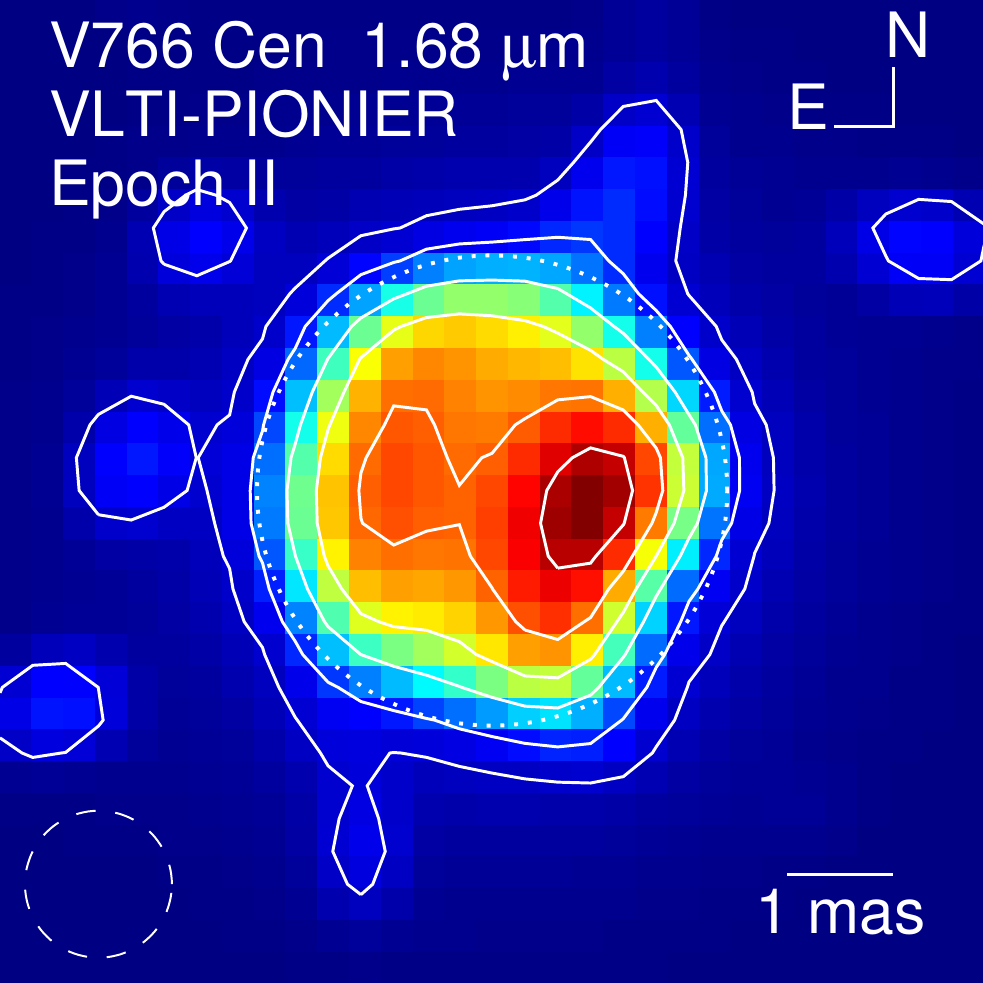}
  \includegraphics[width=0.255\hsize]{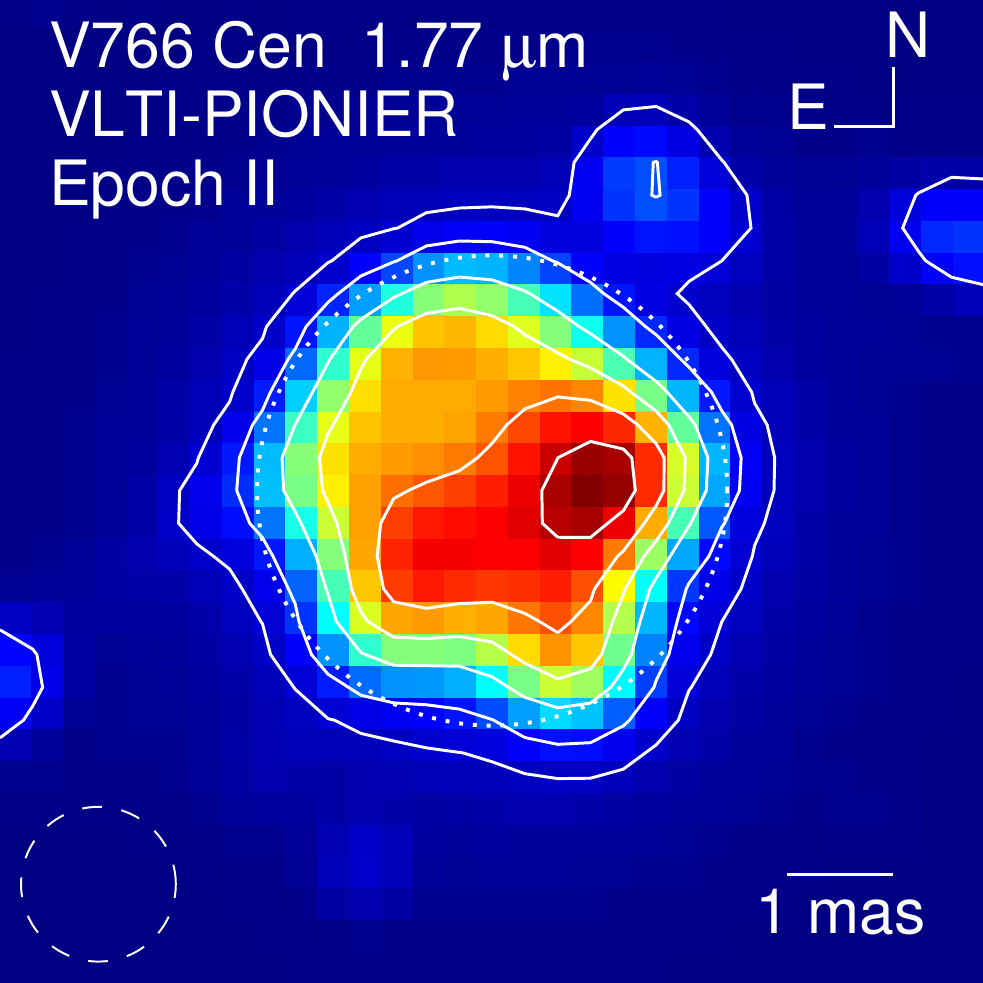}

  \includegraphics[width=0.255\hsize]{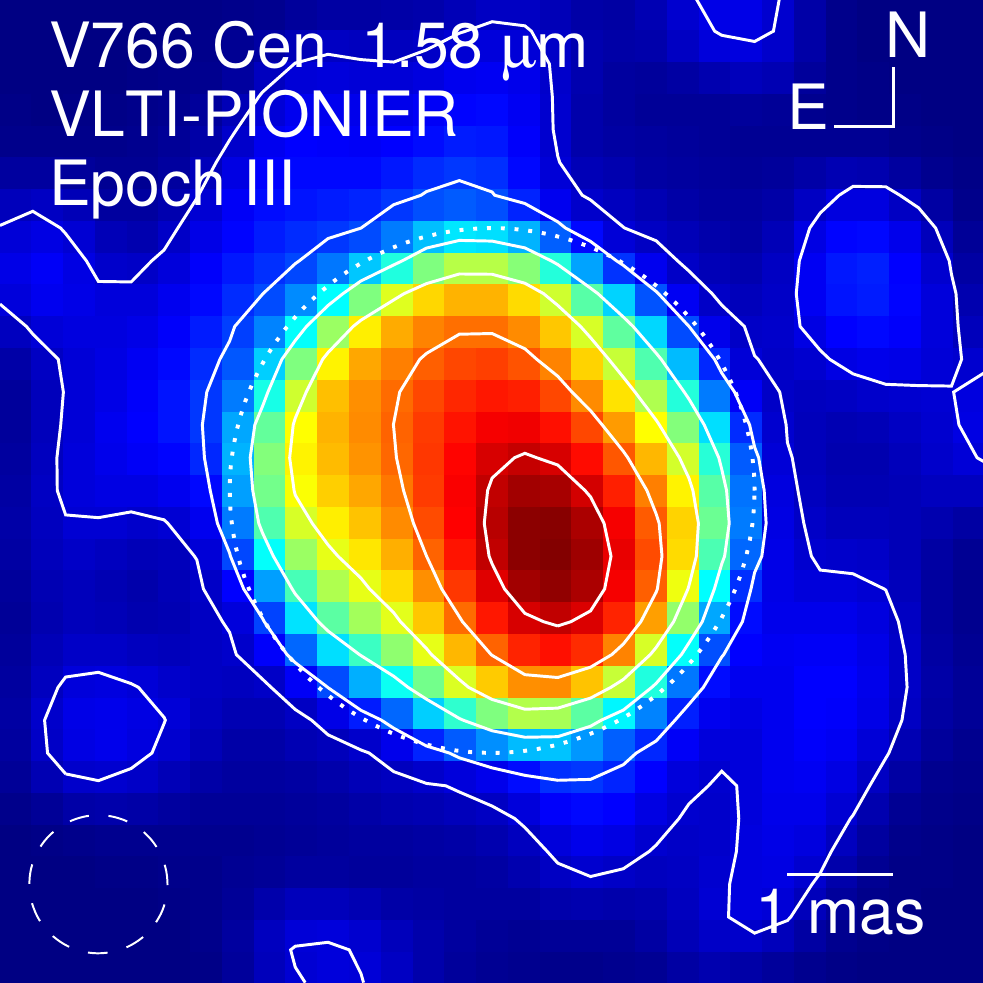}
  \includegraphics[width=0.255\hsize]{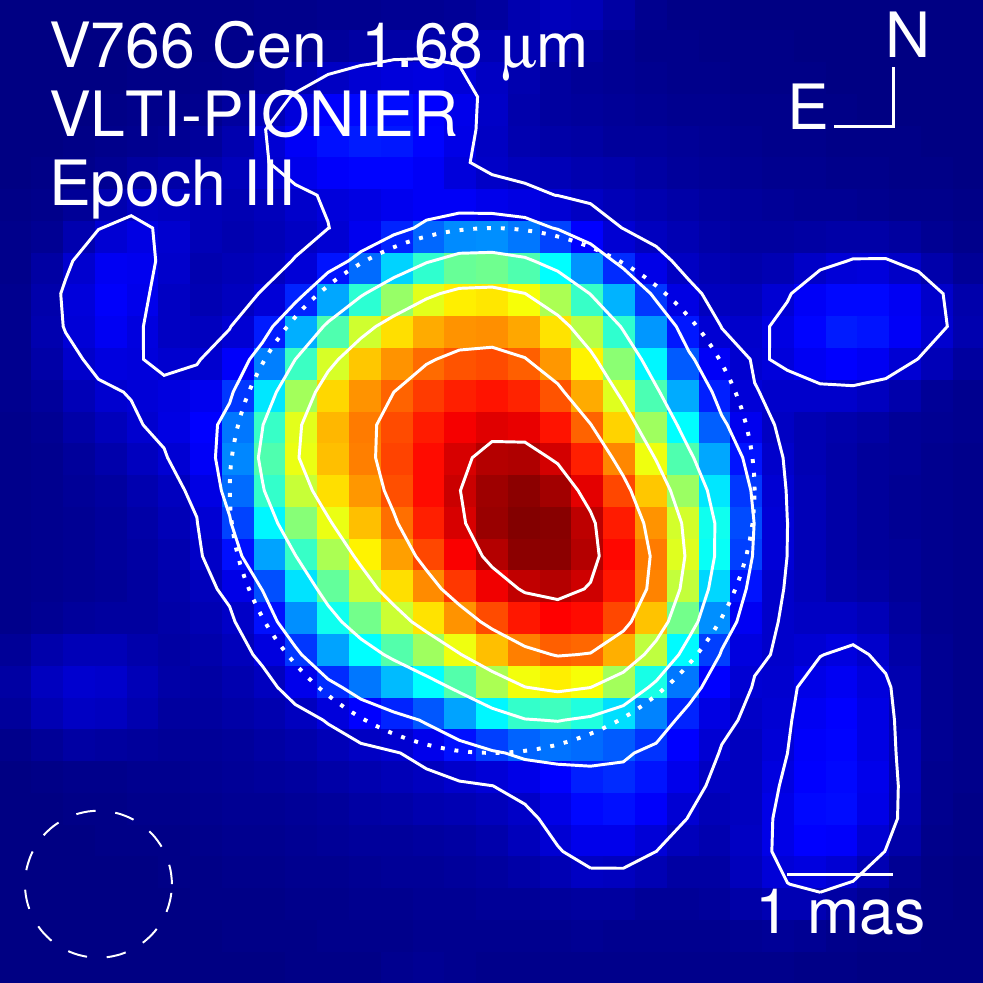}
  \includegraphics[width=0.255\hsize]{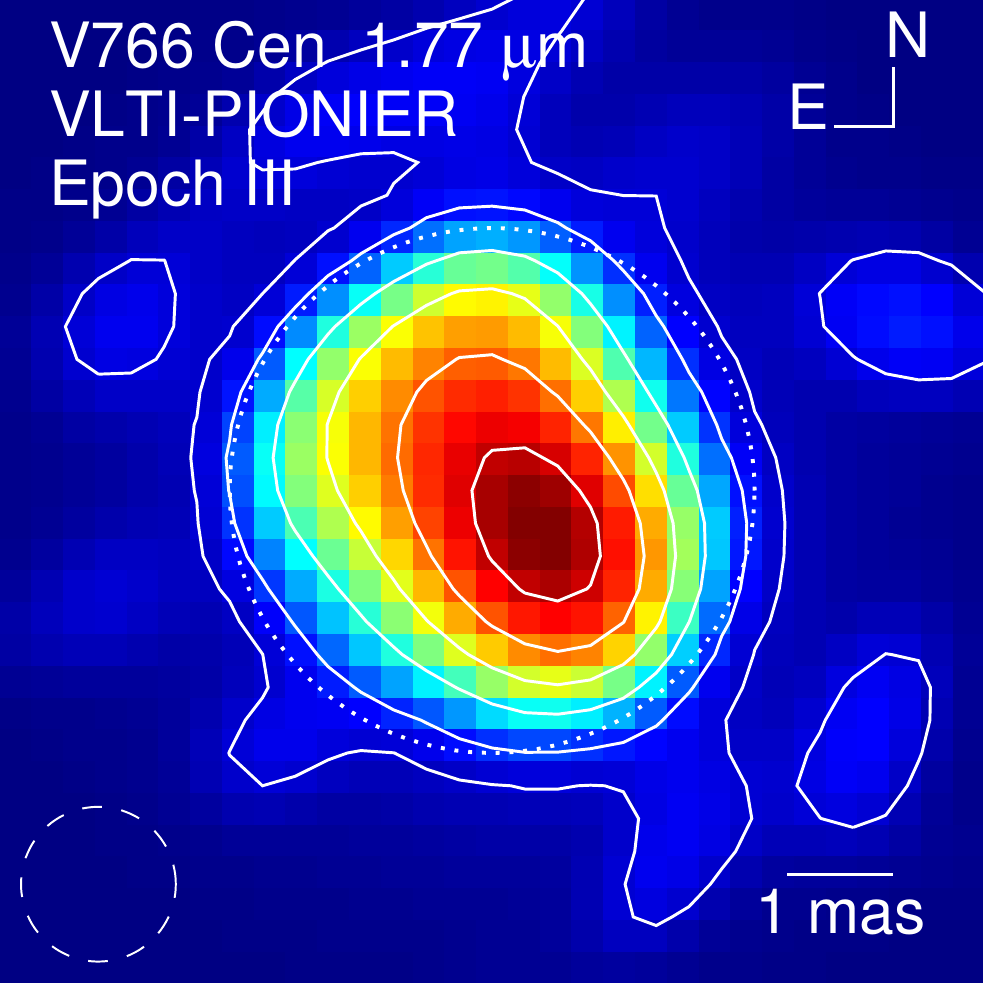}
\caption{
Aperture synthesis images of V766 Cen at 1.58/ 1.59\,$\mu$m (left),
1.68\,$\mu$m (middle), and 1.77\,$\mu$m (right). 
The three rows represent the three epochs.
Contours are drawn at levels of 90\%, 70\%, 50\%, 30\%, 10\%, 3\%
of the peak intensity. The peak intensity is normalized to unity
for each image separately.
The dotted circles indicate our estimated Rosseland angular diameter. 
The dashed circles in the lower left corners indicate the sizes
of the array angular resolution $\lambda/2 B_\mathrm{max}$.
}
\label{fig:images_v766cen}
\end{figure*}
We used the {\tt IRBis} image reconstruction package by
\citet{Hofmann2014} to obtain aperture synthesis
images at each of our three epochs and at each of the
PIONIER spectral channels. The reconstructions were
performed in a similar way as for the carbon AGB star 
R~Scl by \citet{Wittkowski2017b}.

We used 
the best-fit models from Sect.~\ref{sec:analysis} 
as start images. We used a flat prior, and the six available
regularization functions of {\tt IRBis}.
We chose a pixel size of 0.3\,mas,
and we convolved the resulting images with a point spread function (PSF)
of twice the nominal array angular resolution
($\lambda/2B_\mathrm{max} \sim$1.2\,mas).
We used a field of view of 128$\times$128 pixels, corresponding to 
38.4$\times$38.4\,mas, chosen to correspond to twice
the best-fit size of the MOLsphere. As final images, we adopted 
an average of the images obtained with regularization functions 1
(compactness), 3 (smoothness), 4 (edge preservation),
5 (smoothness), and 6 (quadratic Tikhonov), which resulted
in very similar images. Function 2 (maximum entropy) resulted in poorer
reconstructions.

We performed a number of image reconstruction tests including the use of
the reconstruction packages {\tt SQUEEZE} \citep{Baron2010} and
{\tt MiRA} \citep{Thiebaut2008}, further regularization functions,
and different start images. All reconstructions were very similar 
to those obtained with {\tt IRBis}.

Figure~\ref{fig:images_v766cen} shows our reconstructed images 
for the three epochs at three spectral channels with
central wavelengths 1.58/ 1.59\,$\mu$m, 1.68\,$\mu$m, and 1.77\,$\mu$m.
In Fig.~\ref{fig:vis_v766cen}, we over-plot the synthetic squared
visibility amplitudes and closure phases based on the reconstructed images 
to the measured values. The residuals between both of them are also displayed.
The synthetic visibility values based on the reconstructions are in
good agreement with the measured values. There are discrepancies
at small spatial frequencies,
which increase with wavelength. This is a known systematic calibration
effect of PIONIER data
caused by different magnitudes or 
airmass between science and calibrator measurements.
Values of $\chi^2$ for the squared visibility amplitudes range between 
0.4 and 5.3, and for the closure phases between 0.22 and 2.5 for the
different epochs and spectral channels.
The achieved dynamic range varies between about 10 and 20.

The reconstructed images at epoch I show the stellar disk with elongated 
surface features approximately oriented along the East-West direction.
The images at epoch II and epoch III are qualitatively
different to those at epoch I.
They show a dominating narrower single
bright feature.
The feature is located on top of the stellar disk toward its
south-western limb at epoch II and oriented slightly farther toward the 
southern limb at epoch III.
The extended molecular layer or MOLsphere as present in our model
fits from Sect.~\ref{sec:analysis} is not well visible in the 
reconstructed images because it lies just below our achieved
dynamic range.

We estimated the contrast 
$\delta I_\mathrm{rms}/<I>$ \citep[e.g.,][]{Tremblay2013}
of our reconstructed images after dividing them by the best-fit model 
image to correct for the limb-darkening effect, and obtained values 
-- averaged over the spectral channels and regularization functions-- 
of 10\%$\pm$4\%    
for epoch I, 21\%$\pm$6\% for epoch II, and 31\%$\pm$6\% for         
epoch III. The contrasts at epochs II and III are significantly
higher than those at epoch I.

We estimated the angular diameter of the
feature at epochs II and III to 1.7$\pm$0.3\.mas, averaged
over the epochs and spectral channels. This gives a ratio of
0.42$^{+0.35}_{-0.10}$ compared to the Rosseland photospheric radius
of the primary component.

\section{Discussion and conclusions}
\begin{table}
\small
\centering
\caption{Estimated offsets of the companion relative to the primary\protect\label{tab:positions}}
\begin{tabular}{llllll}
\hline\hline
\multicolumn{2}{l}{Epoch}  & JD      & Frac.      & $\Delta$ RA & $\Delta$ Dec \\
         &            & d       & $P_\mathrm{orb}$ & \arcsec     & \arcsec      \\\hline
C    &(2012.18)   & 2455994 & 0                & 1.23       & 0.74        \\
EI   & (2014.17)  & 2456719 & +0.56            & /          & /           \\
EII  & (2016.38) & 2457528 & +1.18            & -0.66      & -0.43       \\
EIII & (2017.23)& 2457839 & +1.41              & -0.89      & -1.38       \\\hline
\end{tabular}
\end{table}
\begin{figure}
\centering
  \includegraphics[width=1.00\hsize]{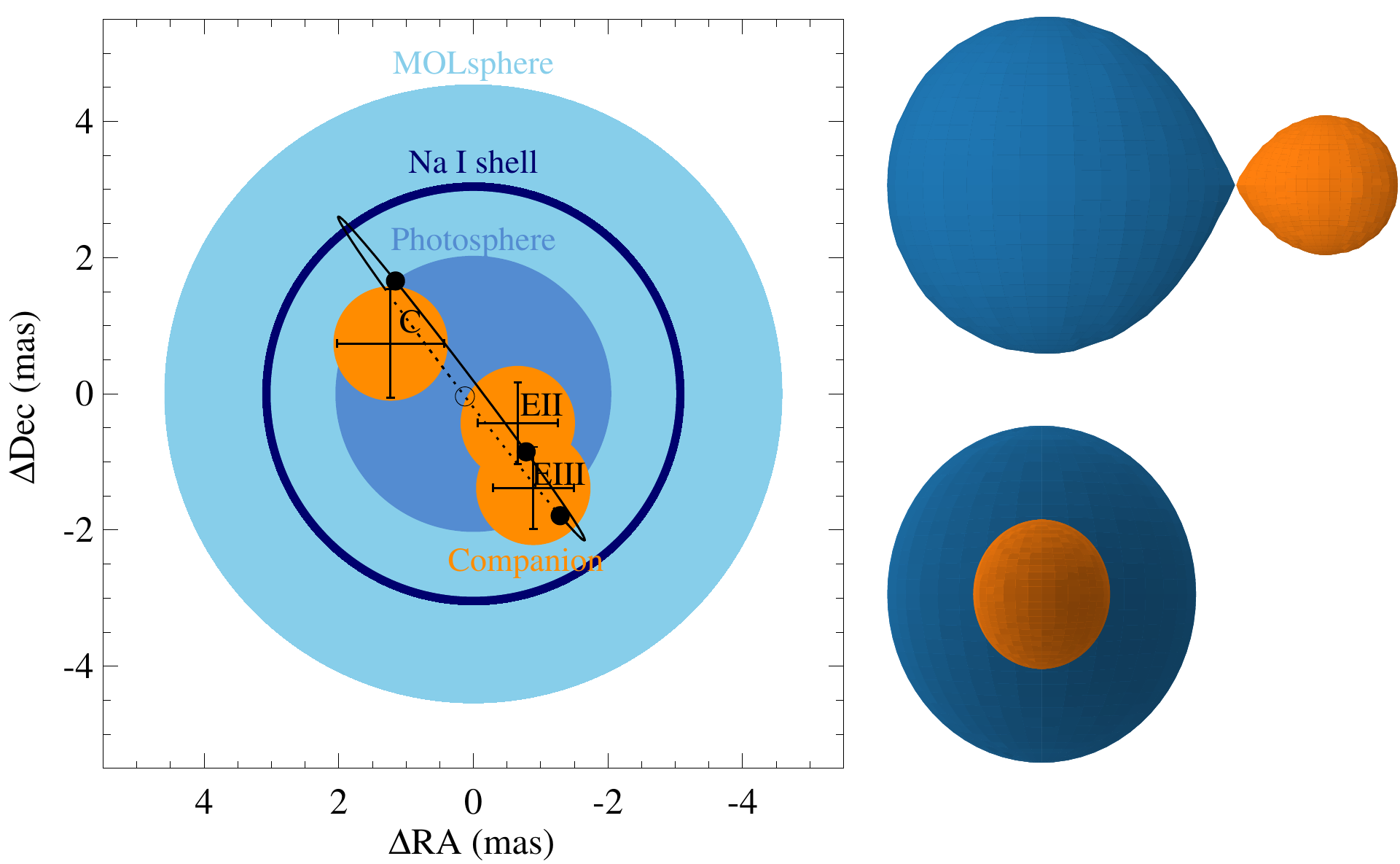}
\caption{Left: Sketch of V766 Cen with the photosphere,
MOLsphere, \ion{Na}{i} shell \citep{Wittkowski2017a},
and the companion at positions from C14 (epoch C)
and our epochs EII and EIII. Also
shown is a plausible Keplerian orbit for a fixed period of 1304\,d and
semi-major axis of 3\,mas. The circles indicate the positions at
our epochs.
Right: Illustration of an in-contact system with our radius ratio,
where the companion is next to the star and we clearly
see the Roche lobes (phase 0.0, top), and in
front of the primary (phase 0.25, bottom). These example phases do
not correspond to the observed intermediate phases in the left panel.}
\label{fig:sketch}
\end{figure}
The images at epoch I are consistent
with predictions by three-dimensional (3D) radiative hydrodynamic (RHD) simulations of
RSGs, such as those shown
by \citet[][Fig. 7]{Chiavassa2010}. As an example, the contrast
of this RHD $H$-band snapshot is 9\%, after convolution to our
spatial resolution and correction for the limb-darkening.
This value is consistent with our observed value of 10\%$\pm$4\%.
We interpret the observed surface features at epoch I by giant
convection cells within the stellar photosphere.

The images at epochs II and III are significantly different to those
at epoch I in terms of their appearance, that is the dominant narrower
feature and their significantly higher contrast. While it may be possible
that the morphology of the convection features has changed from 
epoch I to epochs II and III in this way, the significantly increased 
contrast by a factor of 3 is not consistent with current 3D simulations
such as those mentioned above.
In the following we explore a scenario in which the
images at epochs II and III are dominated by the presence of the close
companion (as suggested by C14) located in front of the
primary and where at epoch I this companion is located behind the primary and
not visible.

For reasons of consistency, we
derived the positions at our epochs with the same fit procedure as in
C14\footnote{There was an error in the sign
convention of the script used for C14, which has now been
corrected.}. Table~\protect\ref{tab:positions} lists the resulting positions.
The best-fit positions agree with our
reconstructed images. We adopt errors of the positions of
half the array angular resolution, that is 0.9\,mas for the 2012
AMBER data and 0.6\,mas for the 2016 and 2017 PIONIER data.
Figure~\ref{fig:sketch} (left) shows a sketch of V766 Cen with the
photospheric disk,
the MOLsphere, the \ion{Na}{i} shell \citep{Wittkowski2017a},
and these companion positions including that of C14.

C14 analyzed available $V$ band light curve data and
available radial velocity data. They could not find conclusive orbital
parameters of the system, meaning that we are not able to compare our positions
to a given orbit. However, they were able to constrain the
orbital period to 1304$\pm$6\,d based on the light curve and
radial velocity data. In order to test whether our companion positions
are consistent with this orbital period and thus with the light curve and
radial velocity data, we
explored Keplerian orbits of V766 Cen and its close companion
using the fixed period of 1304\,d.
With the limitations of the available data, we were not able to
derive any conclusive determination of the orbital parameters.
However, we found an indication that semi-major axes between 2 and 5 mas
with eccentricities as large as 0.5 may produce plausible orbits that
are consistent with our observed angular sizes, our companion positions,
and with the orbital period by C14.
For illustration, Fig.~\ref{fig:sketch} (left) includes an example
of a plausible NE-SW orbit with a semi-major axis of 3\,mas.
This example orbit has a total mass of 108\,$M_\odot$, which is above
the estimates by C14 and \citet{Wittkowski2017a}.
The mass is sensitive to the semi-major axis and goes down to 32\,$M_\odot$
at a semi-major axis of 2\,mas. This may point to smaller angular radii
of the components and a smaller semi-major axis within our error ranges.
Nevertheless, this example illustrates that indeed our positions are
consistent with that of C14 and with their orbital period
based on the light curve and radial velocity data.

In conclusion, we interpret our images in the most likely scenario
of the close eclipsing companion that is located
behind the stellar disk at epoch I and in front of the stellar disk
at epochs II and III. The lower contrast surface features at epoch I
as well as the residual features at epochs II and III are caused by
giant convection cells on the surface of the primary.

Assuming that the system is in contact or is in the common envelope phase,
that is, both stars are filling their Roche lobes, we can derive the
mass ratio of the components directly from the radius ratio by solving
the Roche potential (Fig.~\ref{fig:radiusmass}). For a system in contact,
we found that our radius ratio $R_\mathrm{Comp}/R_\mathrm{Prim}$
of 0.42$^{+0.35}_{-0.10}$ corresponds to a mass ratio
of 0.16$^{+0.40}_{-0.07}$.
This result would only be marginally affected if the system was in the common
envelope phase as it mostly affects the stellar extension along the
orbital plane. Figure~\ref{fig:sketch} (right) illustrates an in-contact system
with such a radius and mass ratio.

Our imaging observations confirm the presence of a close companion
to V766 Cen observed in front of the stellar disk at two epochs. With an
angular diameter of 1.7$\pm$0.3\,mas, corresponding to a radius of
650$\pm$150\,$R_\odot$ and a mass of 2--20\,$M_\odot$, it is most likely
a cool giant or supergiant.
We may be witnessing a system similar to the progenitor system of SN1987\,A,
where a low-mass companion was dissolved during the common
envelope phase when the massive progenitor was a RSG.
\bibliographystyle{aa}
\bibliography{RSG-PIONIER}

\begin{thebibliography}{23}
\expandafter\ifx\csname natexlab\endcsname\relax\def\natexlab#1{#1}\fi

\bibitem[{{Arroyo-Torres} {et~al.}(2013){Arroyo-Torres}, {Wittkowski},
  {Marcaide}, \& {Hauschildt}}]{Arroyo2013}
{Arroyo-Torres}, B., {Wittkowski}, M., {Marcaide}, J.~M., \& {Hauschildt},
  P.~H. 2013, \aap, 554, A76

\bibitem[{{Baron} {et~al.}(2010){Baron}, {Monnier}, \&
  {Kloppenborg}}]{Baron2010}
{Baron}, F., {Monnier}, J.~D., \& {Kloppenborg}, B. 2010, \procspie, 7734,
  77342I

\bibitem[{{Chesneau} {et~al.}(2014){Chesneau}, {Meilland}, {Chapellier},
  {Millour}, {van Genderen}, {Naz{\'e}}, {Smith}, {Spang}, {Smoker}, {Dessart},
  {Kanaan}, {Bendjoya}, {Feast}, {Groh}, {Lobel}, {Nardetto}, {Otero},
  {Oudmaijer}, {Tekola}, {Whitelock}, {Arcos}, {Cur{\'e}}, \&
  {Vanzi}}]{Chesneau2014}
{Chesneau}, O., {Meilland}, A., {Chapellier}, E., {et~al.} 2014, \aap, 563, A71
  (C14)

\bibitem[{{Chiavassa} {et~al.}(2010){Chiavassa}, {Lacour}, {Millour}, {Driebe},
  {Wittkowski}, {Plez}, {Thi{\'e}baut}, {Josselin}, {Freytag}, {Scholz}, \&
  {Haubois}}]{Chiavassa2010}
{Chiavassa}, A., {Lacour}, S., {Millour}, F., {et~al.} 2010, \aap, 511, A51

\bibitem[{{de Jager}(1998)}]{deJager1998}
{de Jager}, C. 1998, \aapr, 8, 145

\bibitem[{{Dessart} {et~al.}(2013){Dessart}, {Hillier}, {Waldman}, \&
  {Livne}}]{Dessart2013}
{Dessart}, L., {Hillier}, D.~J., {Waldman}, R., \& {Livne}, E. 2013, \mnras,
  433, 1745

\bibitem[{{Groh} {et~al.}(2014){Groh}, {Meynet}, {Ekstr{\"o}m}, \&
  {Georgy}}]{Groh2014}
{Groh}, J.~H., {Meynet}, G., {Ekstr{\"o}m}, S., \& {Georgy}, C. 2014, \aap,
  564, A30

\bibitem[{{Groh} {et~al.}(2013){Groh}, {Meynet}, {Georgy}, \&
  {Ekstr{\"o}m}}]{Groh2013}
{Groh}, J.~H., {Meynet}, G., {Georgy}, C., \& {Ekstr{\"o}m}, S. 2013, \aap,
  558, A131

\bibitem[{{Hauschildt} \& {Baron}(1999)}]{Hauschildt1999a}
{Hauschildt}, P.~H. \& {Baron}, E. 1999, Journal of Computational and Applied
  Mathematics, 109, 41

\bibitem[{{Hofmann} {et~al.}(2014){Hofmann}, {Weigelt}, \&
  {Schertl}}]{Hofmann2014}
{Hofmann}, K.-H., {Weigelt}, G., \& {Schertl}, D. 2014, \aap, 565, A48

\bibitem[{{Humphreys} {et~al.}(1971){Humphreys}, {Strecker}, \&
  {Ney}}]{Humphreys1971}
{Humphreys}, R.~M., {Strecker}, D.~W., \& {Ney}, E.~P. 1971, \apjl, 167, L35

\bibitem[{{Lafrasse} {et~al.}(2010){Lafrasse}, {Mella}, {Bonneau}, {Duvert},
  {Delfosse}, {Chesneau}, \& {Chelli}}]{Lafrasse2010}
{Lafrasse}, S., {Mella}, G., {Bonneau}, D., {et~al.} 2010, \procspie, 7734,
  77344E

\bibitem[{{Le Bouquin} {et~al.}(2011){Le Bouquin}, {Berger}, {Lazareff},
  {Zins}, {Haguenauer}, {Jocou}, {Kern}, {Millan-Gabet}, {Traub}, {Absil},
  {Augereau}, {Benisty}, {Blind}, {Bonfils}, {Bourget}, {Delboulbe},
  {Feautrier}, {Germain}, {Gitton}, {Gillier}, {Kiekebusch}, {Kluska},
  {Knudstrup}, {Labeye}, {Lizon}, {Monin}, {Magnard}, {Malbet}, {Maurel},
  {M{\'e}nard}, {Micallef}, {Michaud}, {Montagnier}, {Morel}, {Moulin},
  {Perraut}, {Popovic}, {Rabou}, {Rochat}, {Rojas}, {Roussel}, {Roux},
  {Stadler}, {Stefl}, {Tatulli}, \& {Ventura}}]{LeBouquin2011}
{Le Bouquin}, J.-B., {Berger}, J.-P., {Lazareff}, B., {et~al.} 2011, \aap, 535,
  A67

\bibitem[{{Menon} \& {Heger}(2017)}]{Menon2017}
{Menon}, A. \& {Heger}, A. 2017, ArXiv e-prints 1703.04918

\bibitem[{{Podsiadlowski}(2010)}]{Podsiadlowski2010}
{Podsiadlowski}, P. 2010, \nar, 54, 39

\bibitem[{{Podsiadlowski}(2017)}]{Podsiadlowski2017}
{Podsiadlowski}, P. 2017, ArXiv e-prints 1702.03973

\bibitem[{{Sana} {et~al.}(2012){Sana}, {de Mink}, {de Koter}, {Langer},
  {Evans}, {Gieles}, {Gosset}, {Izzard}, {Le Bouquin}, \&
  {Schneider}}]{Sana2012}
{Sana}, H., {de Mink}, S.~E., {de Koter}, A., {et~al.} 2012, Science, 337, 444

\bibitem[{{Thi{\'e}baut}(2008)}]{Thiebaut2008}
{Thi{\'e}baut}, E. 2008, \procspie, 7013, 70131I

\bibitem[{{Tremblay} {et~al.}(2013){Tremblay}, {Ludwig}, {Freytag}, {Steffen},
  \& {Caffau}}]{Tremblay2013}
{Tremblay}, P.-E., {Ludwig}, H.-G., {Freytag}, B., {Steffen}, M., \& {Caffau},
  E. 2013, \aap, 557, A7

\bibitem[{{Tsuji}(2000)}]{Tsuji2000}
{Tsuji}, T. 2000, \apjl, 540, L99

\bibitem[{{van Genderen}(1992)}]{vanGenderen1992}
{van Genderen}, A.~M. 1992, \aap, 257, 177

\bibitem[{{Wittkowski} {et~al.}(2017{\natexlab{a}}){Wittkowski},
  {Arroyo-Torres}, {Marcaide}, {Abellan}, {Chiavassa}, \&
  {Guirado}}]{Wittkowski2017a}
{Wittkowski}, M., {Arroyo-Torres}, B., {Marcaide}, J.~M., {et~al.}
  2017{\natexlab{a}}, \aap, 597, A9

\bibitem[{{Wittkowski} {et~al.}(2017{\natexlab{b}}){Wittkowski}, {Hofmann},
  {H{\"o}fner}, {Le Bouquin}, {Nowotny}, {Paladini}, {Young}, {Berger},
  {Brunner}, {de Gregorio-Monsalvo}, {Eriksson}, {Hron}, {Humphreys},
  {Lindqvist}, {Maercker}, {Mohamed}, {Olofsson}, {Ramstedt}, \&
  {Weigelt}}]{Wittkowski2017b}
{Wittkowski}, M., {Hofmann}, K.-H., {H{\"o}fner}, S., {et~al.}
  2017{\natexlab{b}}, \aap, 601, A3

\end{thebibliography}
\begin{acknowledgements}
This research has made use of the SIMBAD database,
operated at CDS, France, and of NASA's Astrophysics Data System.
AA acknowledges support from the Spanish MINECO through grant
AYA2015-63939-CO2-1-P, cofunded with FEDER funds.
\end{acknowledgements}
\begin{appendix}
\section{Additional material}
\begin{table}
\centering
\caption{Log of our PIONIER observations.\label{tab:obs_pionier}}
\begin{tabular}{lllrrrr}
\hline\hline
Date           & Epoch & Configuration & \# of obs.\\\hline
2014-02-24     & I     & D0/G1/H0/I1   & 9 \\           
2014-03-06/-07 & I     & A1/G1/J3/K0   & 11 \\          
2016-05-07     & II    & A0/G1/J2/J3   & 6  \\          
2016-05-23/-25 & II    & A0/B2/C1/D0   & 6  \\          
2016-06-01     & II    & D0/G2/J3/K0   & 4  \\          
2016-06-27     & II    & A0/B2/C1/D0   & 2  \\          
2016-07-01     & II    & A0/G1/J2/J3   & 3  \\          
2017-02-24     & III   & A0/B2/C1/D0   & 6 \\           
2017-03-11     & III   & A0/B2/D0/J3   & 2 \\           
2017-03-15/-21 & III   & A0/G1/J2/J3   & 9 \\           
2017-04-22/-23 & III   & D0/G2/J3/K0   & 6 \\           
2017-04-24/-29 & III   & A0/D0/G2/J3   & 4 \\           
\hline
\end{tabular}
\end{table}
\begin{figure*}
\centering
  \includegraphics[width=0.33\hsize]{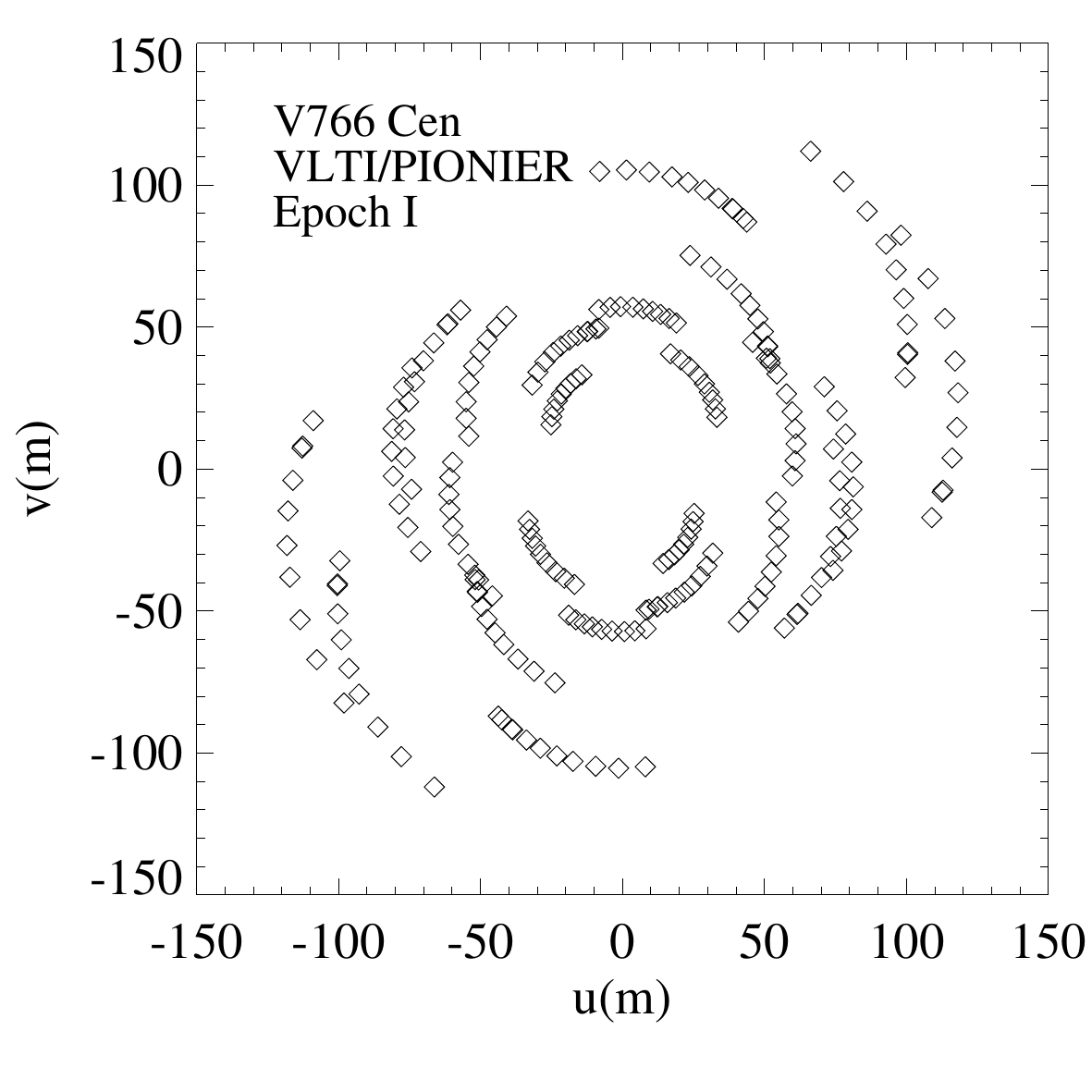}
  \includegraphics[width=0.33\hsize]{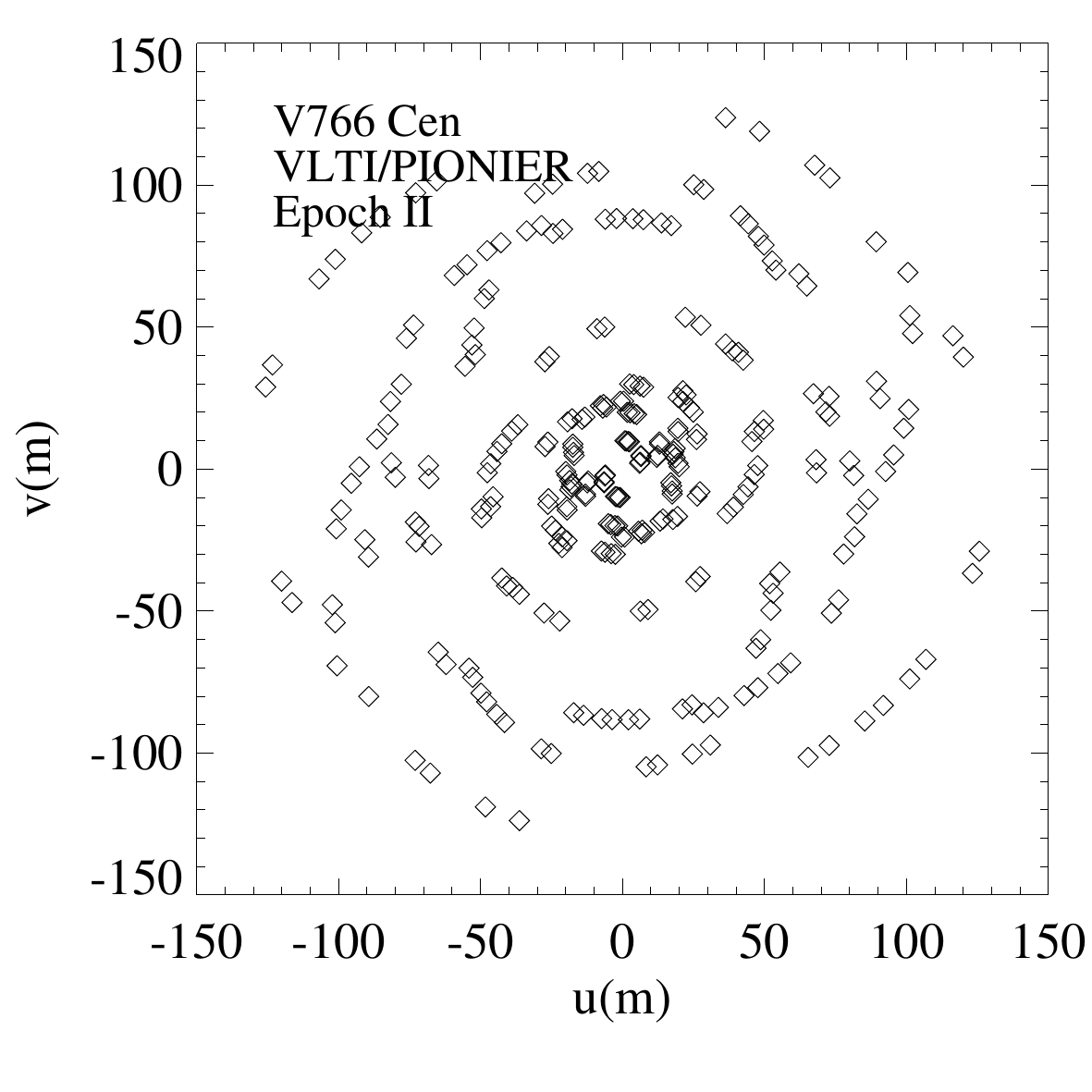}
  \includegraphics[width=0.33\hsize]{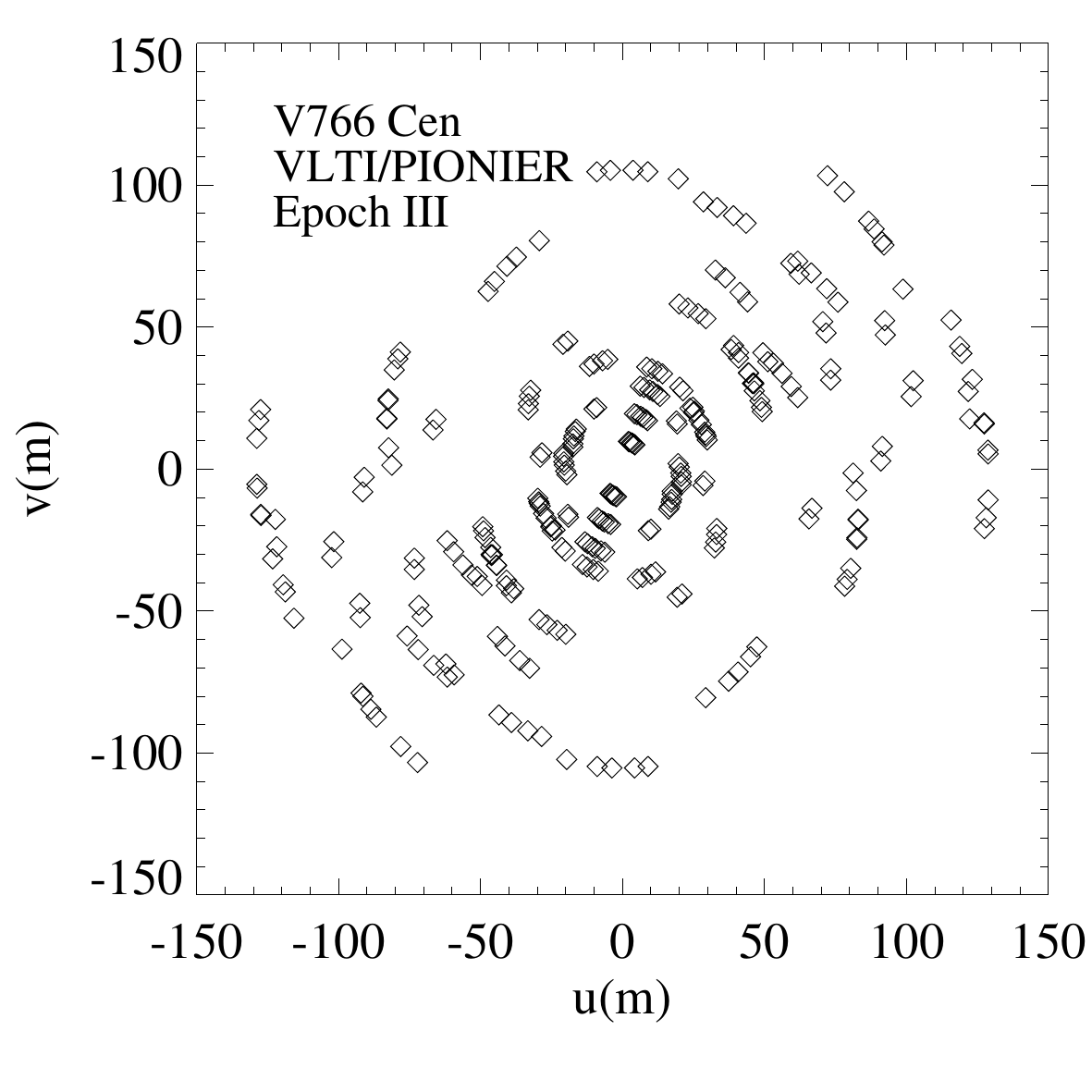}
\caption{The $uv$ coverage obtained for our PIONIER observations of 
V766~Cen at epochs I (left), II (middle), and III (right).}
\label{fig:pionier_uv}
\end{figure*}
\begin{figure*}
\centering
  \includegraphics[width=0.49\hsize]{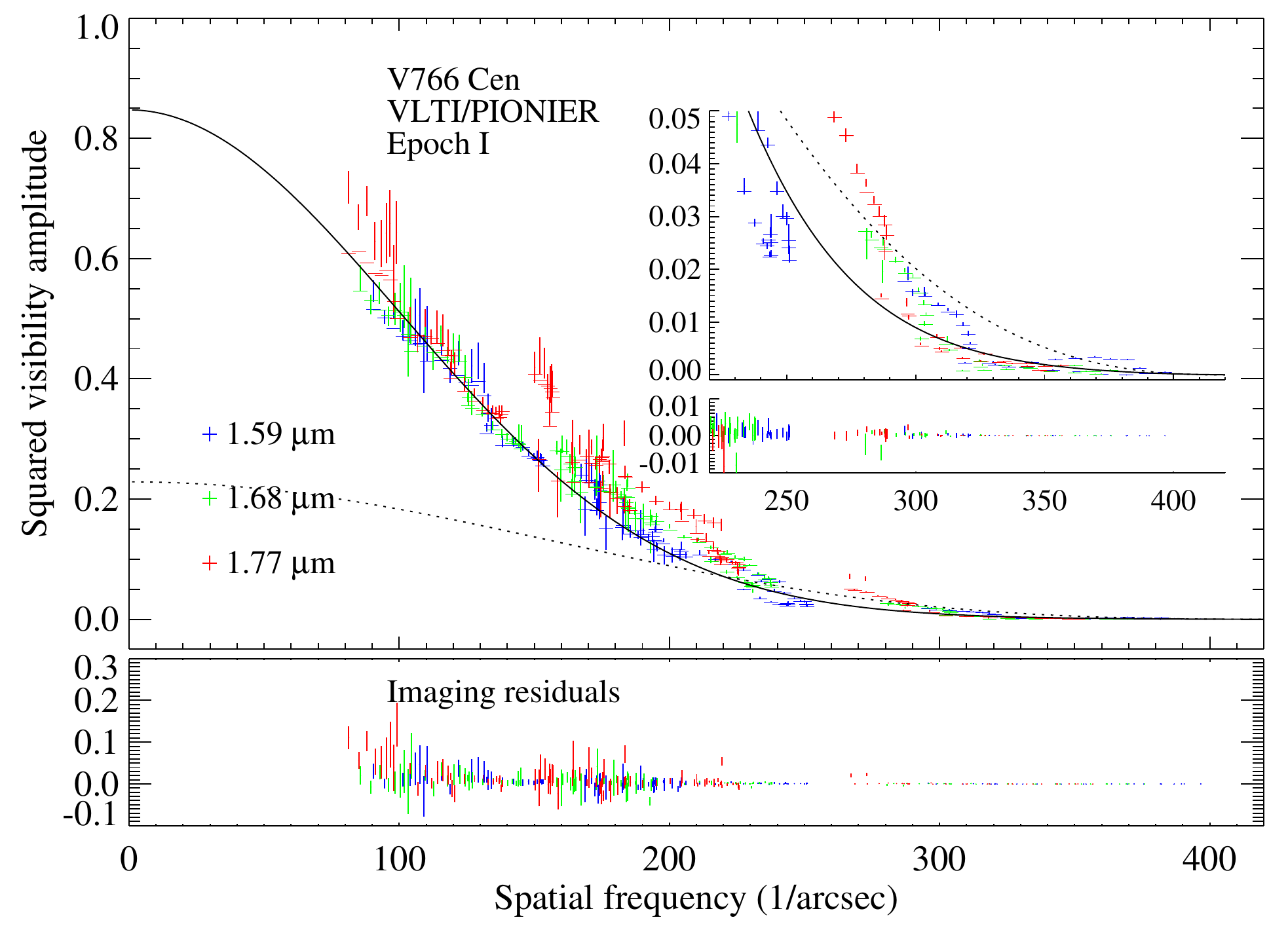}
  \includegraphics[width=0.49\hsize]{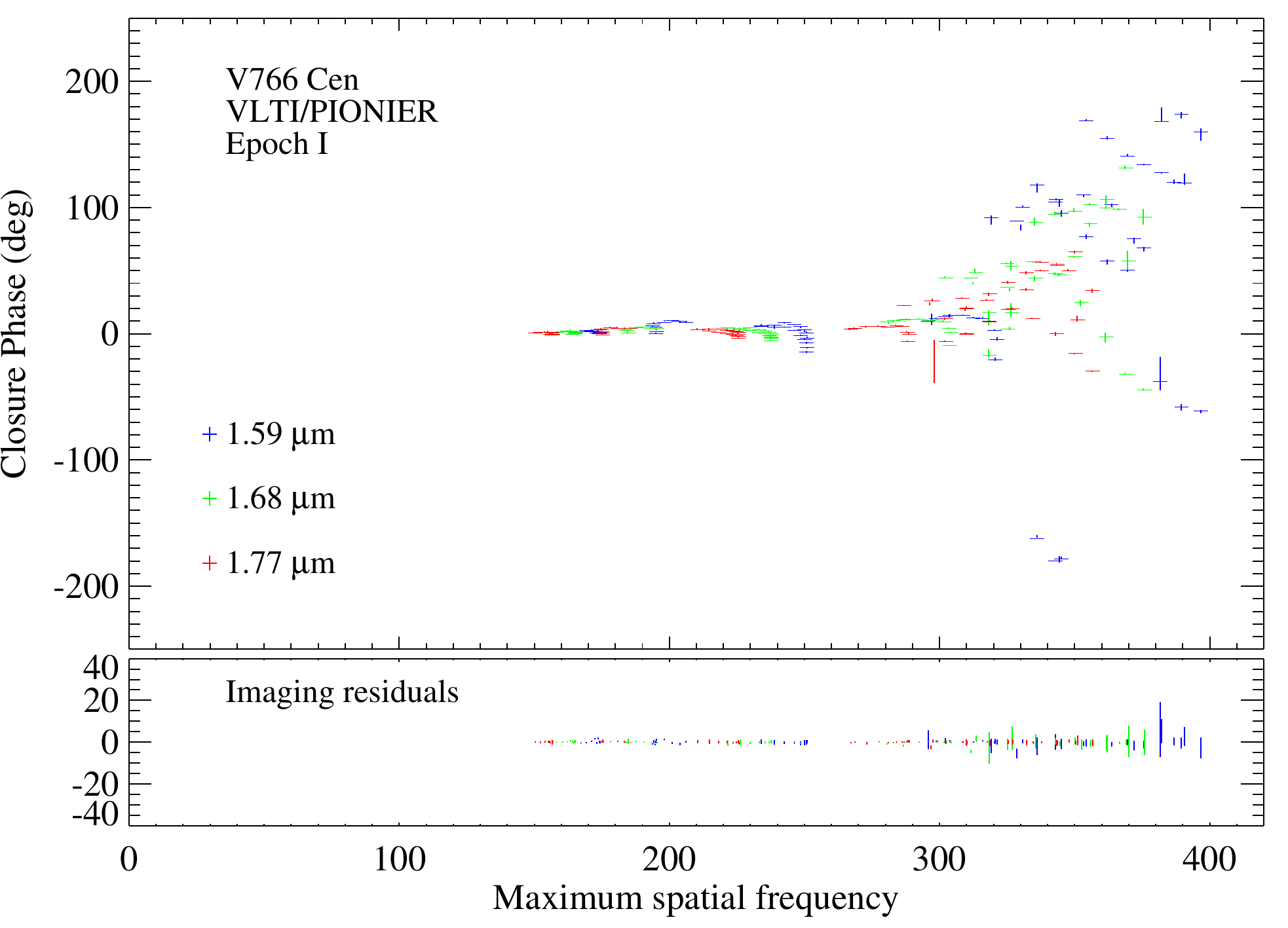}

  \includegraphics[width=0.49\hsize]{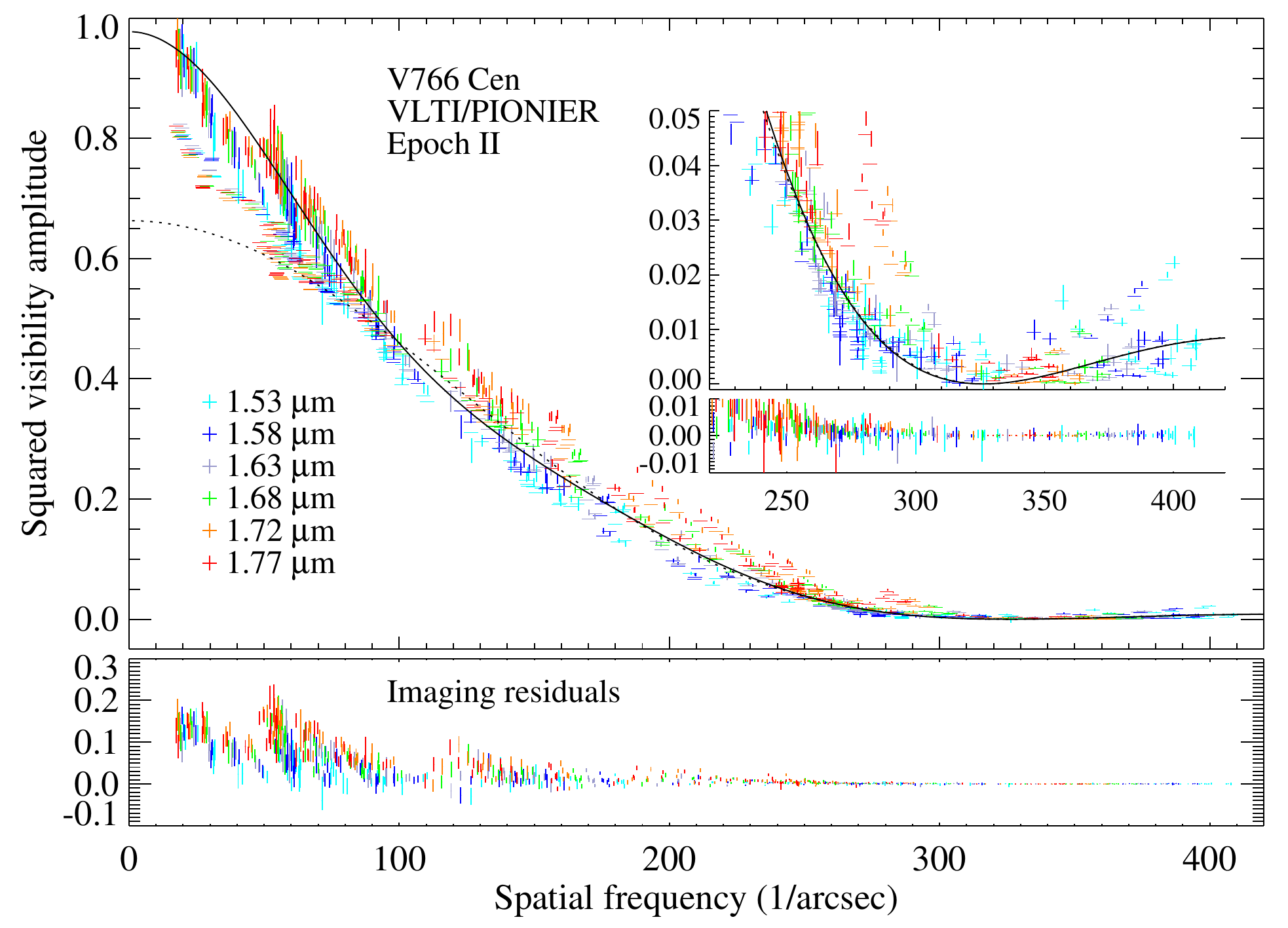}
  \includegraphics[width=0.49\hsize]{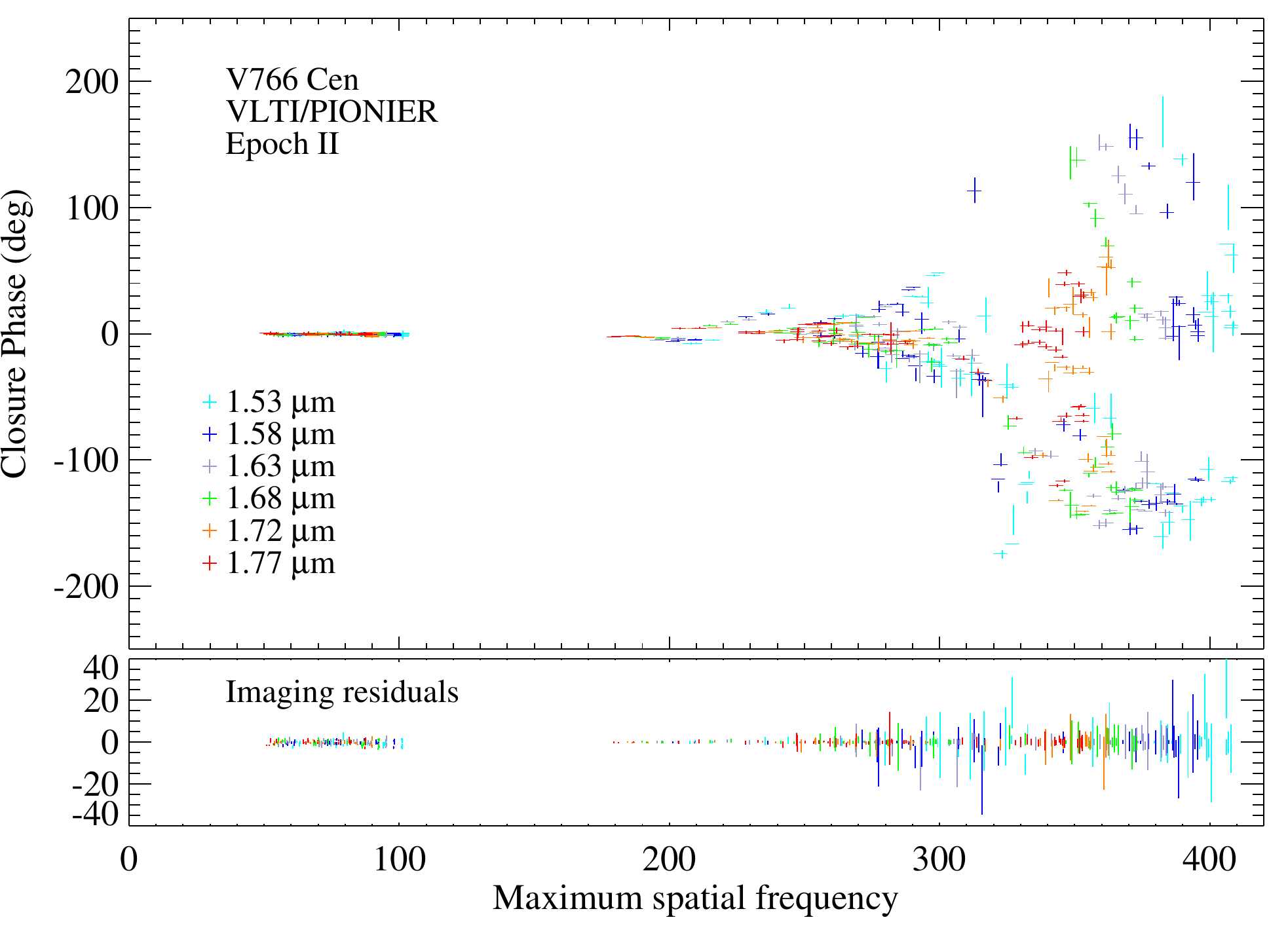}

  \includegraphics[width=0.49\hsize]{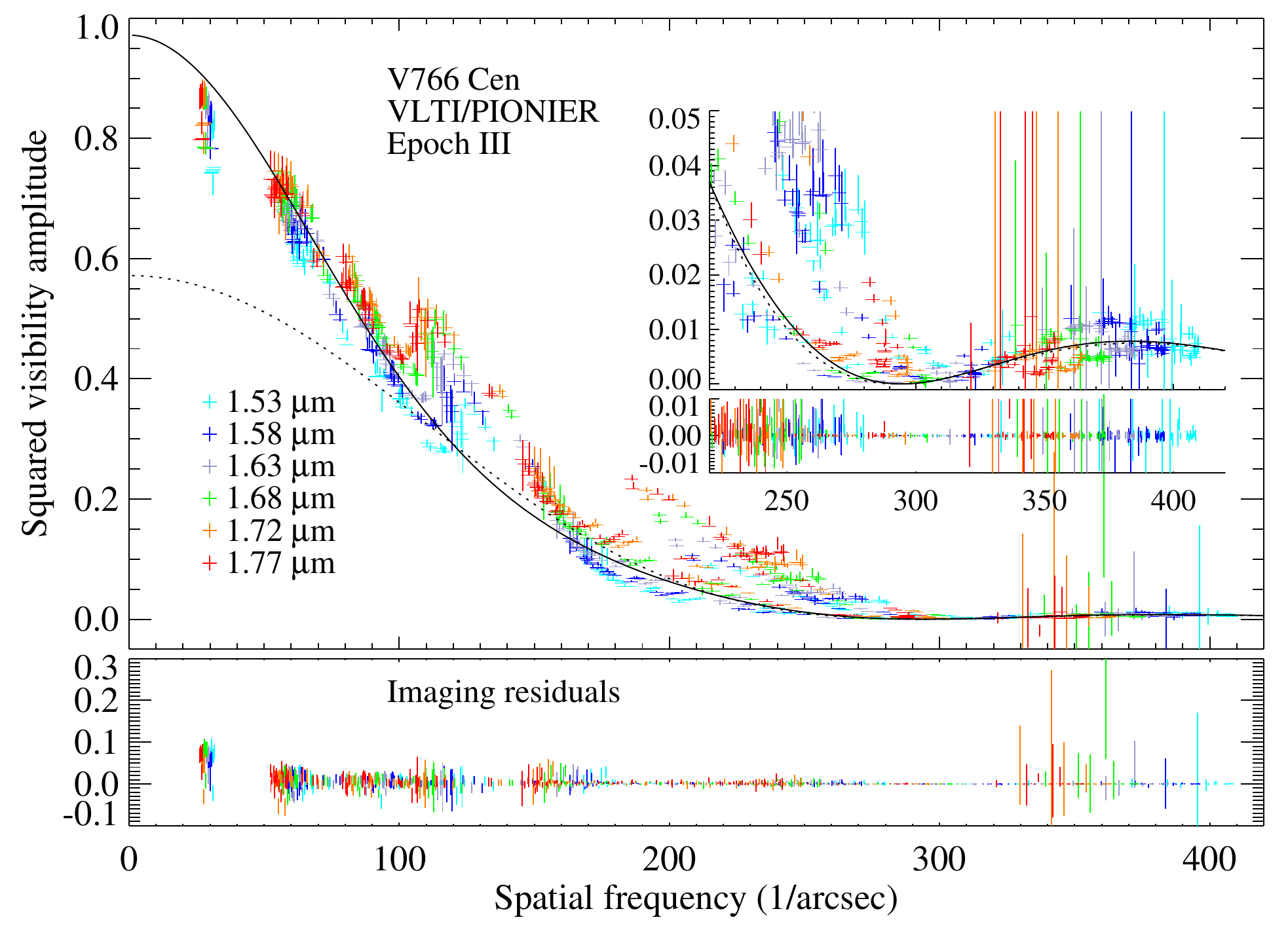}
  \includegraphics[width=0.49\hsize]{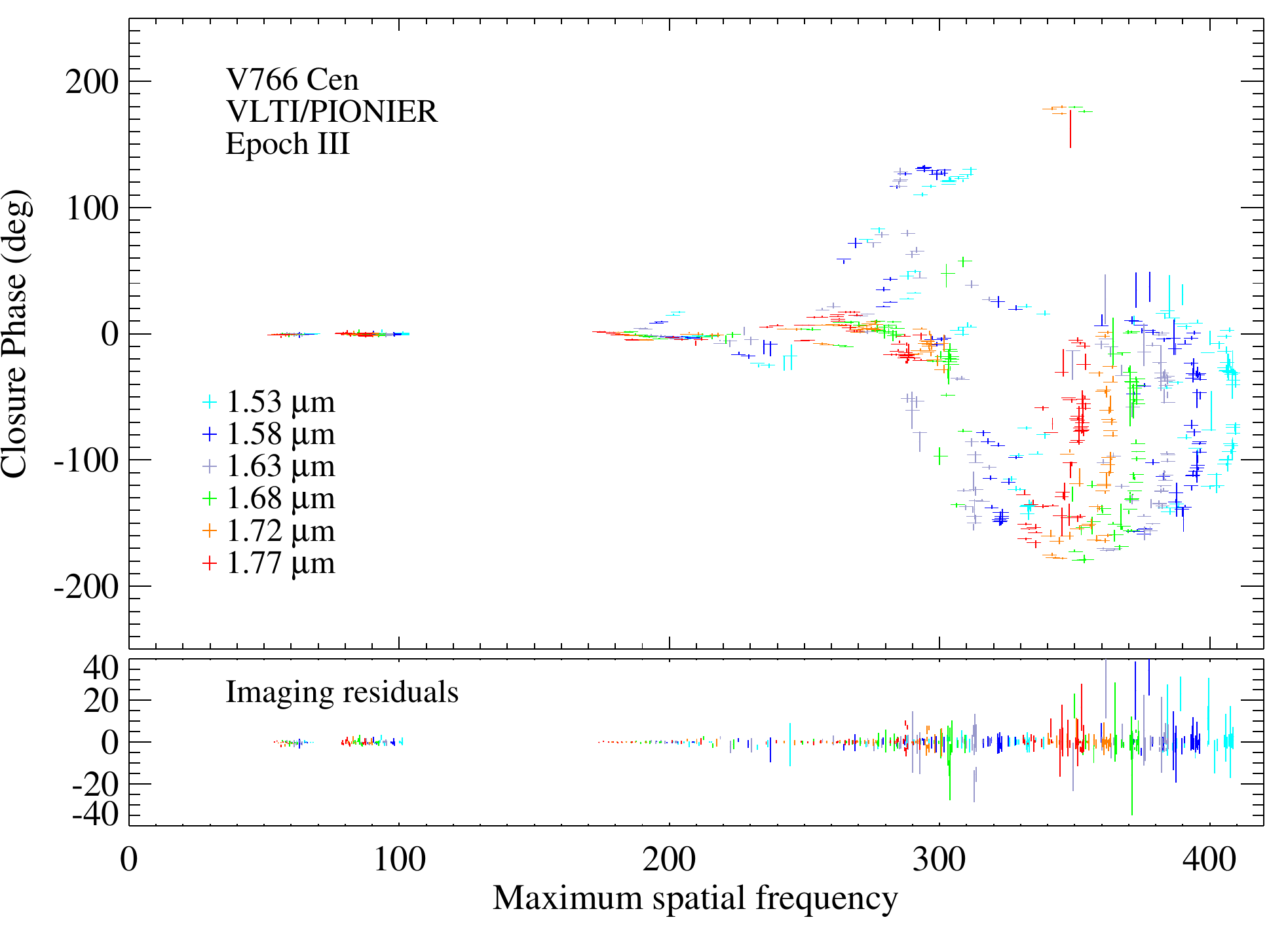}
\caption{Visibility results.
The left panels show the squared visibility amplitudes; 
the inlays enlarge the low values. The right panels show the closure phases. 
The vertical bars indicate the symmetric error bars;
different colors denote different spectral channels. The black solid lines 
denote our visibility model including the stellar photosphere 
represented by a PHOENIX model plus a larger uniform disk 
describing the extended molecular atmosphere. The black dashed lines 
indicate the contribution of the PHOENIX model alone.
The synthetic values based on the reconstructed images are shown by 
horizontal bars. The lower small panels show the residuals between 
observations and reconstructed images.
}
\label{fig:vis_v766cen}
\end{figure*}
\begin{figure}
\centering
\includegraphics[width=1.0\hsize]{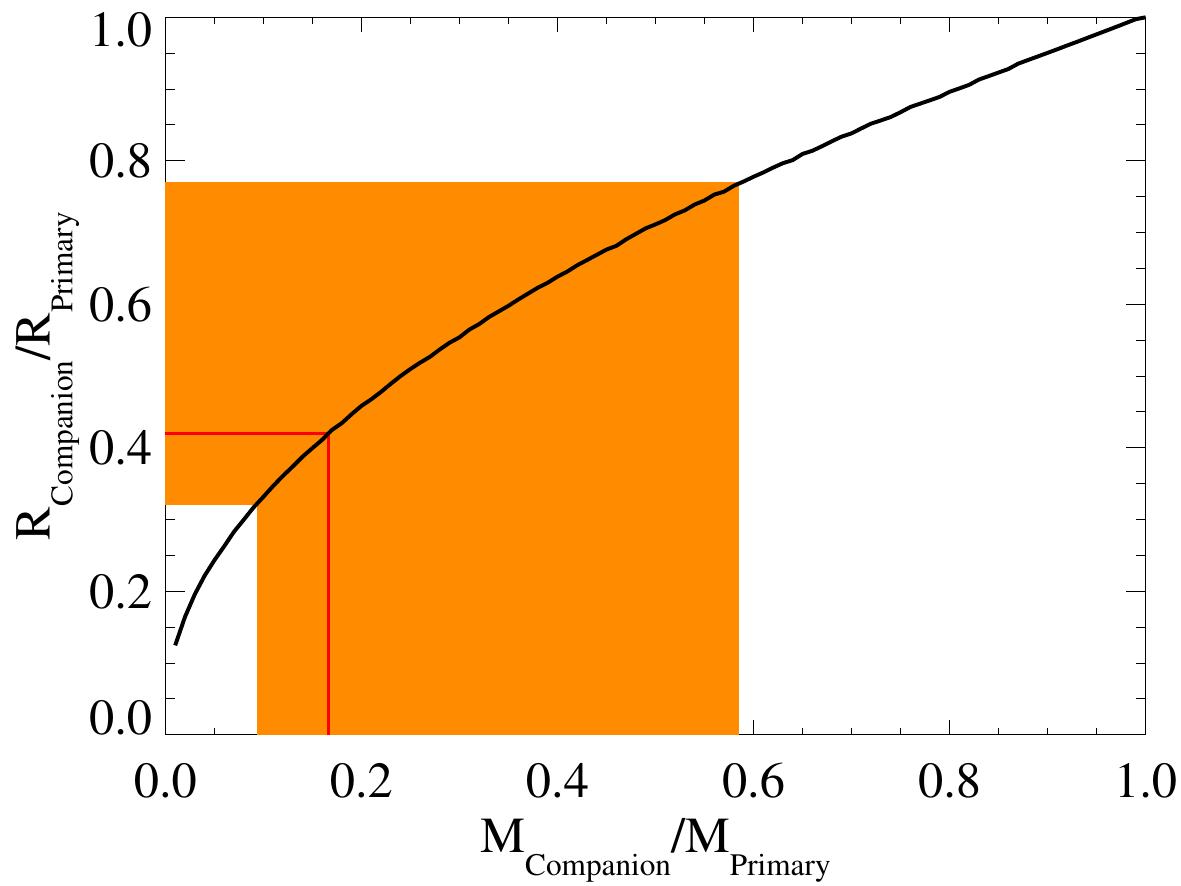}
\caption{The radius ratio $R_\mathrm{Companion}/R_\mathrm{Primary}$
as a function of the mass ratio $M_\mathrm{Companion}/M_\mathrm{Primary}$
of an in-contact system as drawn in \protect Fig. A\ref{fig:sketch} (Left)
The black curve is the theoretical curve from solving the Roche potential.
The red line denotes our observed radius ratio and the corresponding
mass ratio, and the orange area denotes the corresponding uncertainties.}
\label{fig:radiusmass}
\end{figure}
\end{appendix}
\end{document}